\RequirePackage{fix-cm}
\documentclass[epj]{svjour}              
\smartqed  
\usepackage{graphicx}
\usepackage{graphics}
\usepackage{epsfig}
\usepackage[toc,page]{appendix}
\usepackage{bm}
\usepackage{epic,eepic}
\usepackage{epsfig}
\usepackage{pifont}
\usepackage{subfigure}
\usepackage{caption}
\usepackage{amsmath,amssymb,amsfonts,mathrsfs,dsfont,mathtools}
\usepackage{url}
\usepackage{color}
\usepackage{soul}

\newcommand{\be}{\begin{equation}}
\newcommand{\ee}{\end{equation}}

\usepackage{float}

\usepackage{epstopdf}
\usepackage{mathptmx}
\usepackage{latexsym}


\begin{document}
\title{Sliding droplets of Xanthan solutions: a joint experimental and numerical study}
\author{Silvia Varagnolo\inst{1} \and Giampaolo Mistura\inst{1} \and Matteo Pierno\inst{1}  \and Mauro Sbragaglia\inst{2}  
}                     
%
%
\institute{Dipartimento di Fisica e Astronomia `G. Galilei' and CNISM, Universit\`a di Padova, Via Marzolo, 8 - 35131  Padova  PD (Italy) \and Department of Physics and INFN, University of ``Tor Vergata'', Via della Ricerca Scientifica 1, 00133 Rome, Italy}
\date{Received: date / Revised version: date}
%
\abstract{
We have investigated the sliding of droplets made of solutions of Xanthan, a stiff rodlike polysaccharide exhibiting a non-newtonian behavior, notably characterized by a shear-rate dependence of the viscosity. These experimental results are quantitatively compared with those of newtonian fluids (water). The impact of the non-newtonian behavior on the sliding process was shown through the relation between the average dimensionless velocity (i.e. the Capillary number) and the dimensionless volume forces (i.e. the Bond number). To this aim, it is needed to define operative strategies to compute the Capillary number for the shear thinning fluids and compare with the corresponding newtonian case. Results from experiments were complemented with lattice Boltzmann numerical simulations of sliding droplets, aimed to disentangle the influence that non-newtonian flow properties have on the sliding. 
} 
\maketitle

\section{Introduction}

Controlling and predicting the properties of spreading and moving droplets is a major scientific challenge, relevant for an ample variety of applications, particularly in droplet-based microfluidics \cite{cira15,mannetje14,jebrail11,sinz12,Gau99,lagubeau2011,piroird13,sartori15}. From the point of view of fundamental research \cite{Kistler97,DeGennes85,Furmidge62,Huh71,Buehrle02,seemann05,Podgorski01,Kimetal02,Rio05,PRLnoi,PREnoi,Langmuir}, the spreading and motion of a droplet on a solid surface has received much attention because of the singularity that occurs at the contact line \cite{Huh71}, where the competition between capillary forces and viscous dissipation results in a singular problem \cite{Podgorski01,Kimetal02,Rio05}. Close to the contact line, the flow is confined to a wedge-like region whose shape is largely independent of the macroscopic geometry. An added richness is brought by the complex non-newtonian behavior of the bulk phases, e.g. concentrated polymer solutions, which often exhibit both shear-thinning and normal stress effects~\cite{bird87,Herrchen97,Lindner03,Arratia}. Deformation rates imposed on the fluid may indeed induce non-newtonian behavior within the wedge-like region~\cite{WeiGaroff09,WeiGaroff07}. Shear thinning was even proposed as mitigating the moving contact line singularity: if the viscosity of the liquid decreases monotonically with increasing shear stress, then there will still be an infinite shear stress, but the stress singularity will be integrable, yielding a finite force at the substrate~\cite{Weidner94,Ansini02,Carre02}. Some experimental studies analyzed the effects of non-newtonian rheology on the problem of spreading~\cite{Carre97,Carre00,Rafai04,Rafai05}. The spreading behavior was found not to deviate strongly from that of the newtonian fluids~\cite{Tanner79}. The experiments were also compared with the predictions of lubrication theory of power-law fluids \cite{Baudaud,RoHomsy}. Other studies on non-newtonian effects focused on coating flows \cite{Borkar94,Bajaj08,Abedijaberi11,Fraysse94} and moving contact lines \cite{WeiGaroff09,WeiGaroff07,Seevaratnam05,Seevaratnam07,Yueetal12}, both experimentally \cite{WeiGaroff09,WeiGaroff07,Seevaratnam05,Seevaratnam07} and numerically/theoretically \cite{Yueetal12,HanKim13,HanKim14}. Garoff and co-workers \cite{WeiGaroff09,WeiGaroff07,Seevaratnam05,Seevaratnam07} conducted a series of experiments on dynamic wetting: by comparing the steady interface shape, they observed that the non-newtonian effect was confined to the close vicinity of the contact line. In this inner region, the viscous bending of the interface is reduced by shear thinning and enhanced by viscoelasticity, a fact that was also confirmed by numerical simulations \cite{Yueetal12}.\\
In this paper we report on a joint experimental/numerical study aimed to investigate the sliding of droplets made by polymeric solutions exhibiting non-newtonian properties. The corresponding newtonian problem has been studied in an ample variety of situations \cite{Furmidge62,Huh71,Podgorski01,Kimetal02,Rio05,PRLnoi,PREnoi,Langmuir,kusumaatmaja06,kusumaatmaja07,Stone04}, but the effects of viscoelasticity have been rarely investigated \cite{Kieweg13,Ahmed13,SchwartzOral08,Perazzo03,Perazzo05,Saccomandi12}, especially from the experimental point of view. As far as we know, only droplets of polystyrene solutions have been explored experimentally on hydrophilic substrates, where sliding is often affected by the presence of pearling~\cite{Morita09}. Most non-newtonian fluids exhibit various non-newtonian properties: notably, concentrated polymer solutions are characterized by both shear-thinning and normal stress effects~\cite{Carre97,Carre00,Rafai04,Rafai05}. With the aim of concentrating on the shear-thinning flow properties, we will consider solutions of Xanthan, a stiff rodlike polysaccharide. Xanthan solutions are chosen because, when compared to other flexible polymer solutions  \cite{Rafai04}, are reported to have a strong shear rate dependence of the viscosity, and smaller elastic effects (i.e. normal stresses)~\cite{Carre97,Carre00,Rafai04,Rafai05}. Operatively, the impact of a shear-dependent viscosity on the sliding process is analyzed through the relation between the average dimensionless velocity (i.e. the Capillary number) and the dimensionless volume forces (i.e. the Bond number). Results from experiments are complemented with numerical simulations of sliding droplets with the lattice Boltzmann models (LBM) \cite{PREnoi,Langmuir}. Numerical simulations provide an extremely versatile tool to investigate the effects of non-newtonian rheology and to elucidate which deviations from the newtonian result could be attributed to a specific non-newtonian property.\\
The paper is organized as follows: section \ref{sec:Experiments} deals with the experimental details, while the corresponding numerical results are described in section \ref{sec:Simulations}. Conclusions will follow in section \ref{sec:conclusion}.

\section{Experiments}\label{sec:Experiments}

We have investigated shear thinning solutions of Xanthan (molecular weight, $M_{w}\sim 10^{6}$ g $mol^{-1}$ , Sigma Aldrich), a stiff rodlike polysaccharide. Solutions have concentrations ranging in the dilute or semi-dilute regime \cite{Rafai04,Rafai05,Callaghan00} as listed in Table \ref{viscosity_table}. These non-newtonian solutions are mainly characterized by a shear-dependent viscosity, but also normal stress effects have been reported, particularly from high enough ($\sim$ 1000 ppm) concentrations \cite{Rafai04,Macosko78,Helmreich95,BonnMeunier97,Zirnsak99,Lopez02,Stokes11,Gupta86,Jones87,Binnington87,Chai90}. While studies on the shear viscosity of Xanthan gum solutions are well established, it seems from the literature that a smaller number of studies have looked systematically and quantitatively at the first normal stress difference $N_1$ \cite{Rafai04,Macosko78,Helmreich95,BonnMeunier97,Zirnsak99,Lopez02,Stokes11,Gupta86,Jones87,Binnington87,Chai90,Bewersdorff88}. Specifically, some studies focused on the scaling of $N_1$ {\it vs.} $\dot{\gamma}$ for various polymer concentrations. Different scaling laws were reported \cite{Macosko78,Stokes11,Gupta86,Jones87,Binnington87}, depending also on the range of shear rates used \cite{Macosko78,Binnington87,Chai90}. In other works \cite{Rafai04}, where the behavior of Xanthan solutions was compared to other polymeric solutions (e.g. polyacrylamide), normal stresses were claimed to be negligible if compared to the strong shear rate dependence of the viscosity.


\begin{table}[!htb]
\centering 
\begin{tabular}{|c|c|c|c|}
\hline
Liquid & concentration  & $a$ & $b$ \\ 
& (ppm w/w) & (Pa $s^{1-b}$) &  \\ \hline
Xanthan/water & 400 & 0.033  & 0.41 \\ 
Xanthan/water & 800 & 0.076  & 0.51 \\ 
Xanthan/water & 1500 &  0.31  & 0.61 \\ 
Xanthan/water & 2500 &  0.99 & 0.70 \\ \hline
\end{tabular}%
\caption{Viscosity parameters for the non-newtonian Xanthan solutions: $a$ and $b$ are the parameters fitting the rheological data described by the power law $\eta_{xan}$($\dot{\gamma}$)=$a$$\dot{\gamma}^{-b}$ and plotted in Figure \ref{fig:ReoXanthan}.}
\label{viscosity_table}
\end{table}


\begin{figure}[t!]
\includegraphics[scale=1]{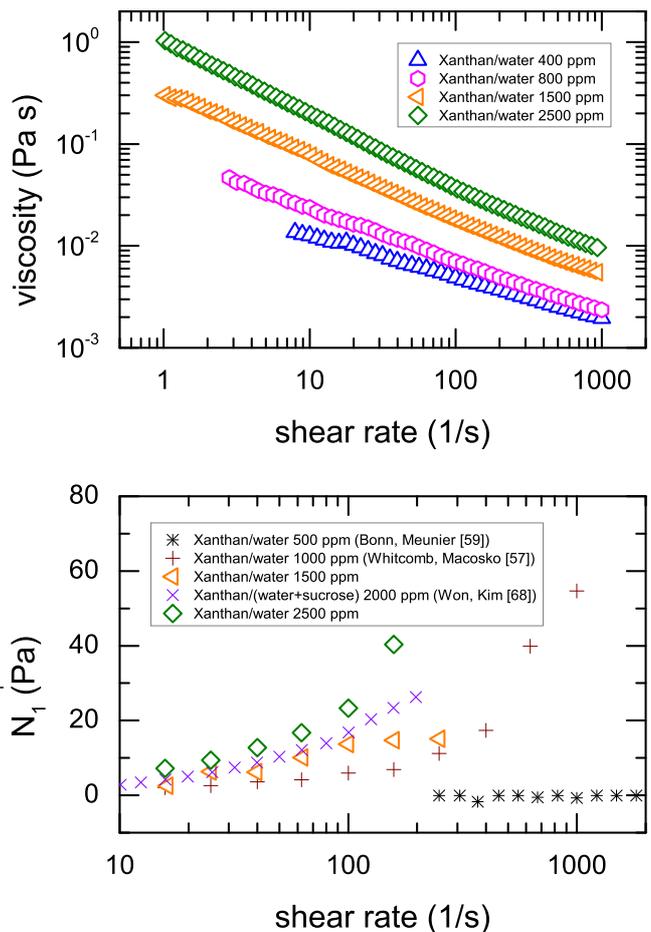}
\caption{Rheological measurements performed with a parallel plate rheometer. Top Panel: viscosity characterization for Xanthan solutions with different concentrations. The viscosity varies with the shear rate $\dot{\gamma}$ according to the power law  $\eta_{xan}(\dot{\gamma})=a \dot{\gamma}^{-b}$, with $a$ and $b$ dependent on the concentration, in agreement with \cite{Rafai04}. Bottom Panel: $N_1$ measurement for 1500 and 2500 ppm Xanthan solutions in comparison with literature data \cite{BonnMeunier97,Macosko78,Won04}. \label{fig:ReoXanthan}}
\end{figure}

We characterized the rheological behavior of our solutions through a parallel plate rheometer (Advanced Rheometric Expansion System ARES, TA instruments) equipped with two 50 mm diameter disks. As shown in the top panel of Figure \ref{fig:ReoXanthan}, the viscosity varies with the shear rate $\dot{\gamma}$ according to the power law
$$
\eta_{xan}(\dot{\gamma})=a \dot{\gamma}^{-b}
$$ 
with $a$ and $b$ dependent on the concentration, as listed in Table \ref{viscosity_table} (see also \cite{Rafai04}). The bottom panel of Figure \ref{fig:ReoXanthan} reports the evaluation of the first normal stress difference $N_1$ achievable by our rheometer, i.e. the values about the 1500 and 2500 ppm solution: values related to lower concentrations were found to be smaller, actually negligible below 1000 ppm \cite{BonnMeunier97}, and also a bit more scattered than data at higher concentrations because near to the sensitivity limit of the instrument. Our measurement indicates the presence of a measurable $N_1$ in agreement with \cite{Macosko78,Won04}, which investigate the rheology of Xanthan solutions of similar concentrations (see Figure \ref{fig:ReoXanthan}). \\
The substrate used in all experiments was a homogeneous, polycarbonate (PC) plate, whose wettability properties have been determined through the sessile droplet method. The sample is characterized by a static contact angle $\theta $ close to $90^{\circ }$ and dynamic contact angles similar for all solutions, as reported in Table \ref{angoli_contatto}. 


\begin{table*}[!htb]
\centering 
\begin{tabular}{|c|c|c|c|c|c|}
\hline
Liquid & concentration & $\theta$ & $\theta_{A}$ & $\theta_{R}$ & $\Delta$$%
\theta$ \\ 
& (ppm w/w) & (degrees) & (degrees) & (degrees) & (degrees)\\ \hline
water &  & 81$\pm$3 & 91$\pm$2 & 66$\pm$3 & 25$\pm$4 \\ 
Xanthan/water & 400 & 86$\pm$2 & 89$\pm$2 & 64$\pm$2 & 24$\pm$3\\ 
Xanthan/water & 800 & 87$\pm$2 & 92$\pm$2 & 67$\pm$3 & 26$\pm$3  \\ 
Xanthan/water & 1500 & 89$\pm$1 & 91$\pm$1 & 66$\pm$1 & 25$\pm$2 \\ 
Xanthan/water & 2500 & 87$\pm$3 & 89$\pm$1 & 63$\pm$2 & 25$\pm$2 \\ \hline
\end{tabular}%
\caption{Static ($\protect\theta$), advancing ($\protect\theta_{A}$),
receding ($\protect\theta_{R}$) contact angles and contact angle hysteresis (%
$\Delta$$\protect\theta$) of all the investigated solutions on a homogeneous
PC sample. }
\label{angoli_contatto}
\end{table*}

\begin{figure}[!htb]
\begin{center}
\includegraphics[scale=0.25]{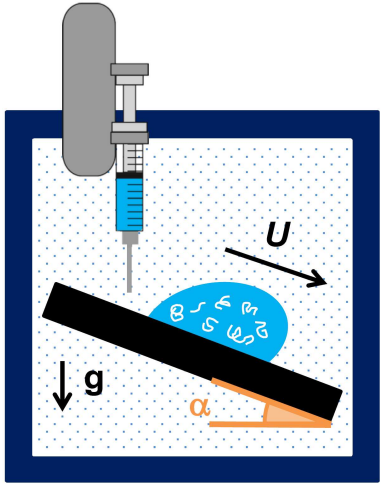}
\caption{Sketch of the experiment: a droplet containing Xanthan at a given concentration is deposited on the polycarbonate surface, inclined by the angle $\alpha$, through a vertical syringe pump dosing $\simeq30$ $\mu \text{L}$ with an accuracy of about 3\%. The droplet is illuminated by a square 5$\times$5 cm$^2$ white LED backlight. The drop velocity $U$ is measured by acquiring the whole sliding sequence,  $g$ being the gravity acceleration. \label{sketch}}
\end{center}
\end{figure}

Sliding measurements of 30 $\mu \text{L}$ droplets have been performed by depositing the desired volume on the already inclined sample (see figure \ref{sketch}) and recording droplet motion with a CMOS camera (MV-D1024E Camera Link), whose maximum frame rate is almost 120 fps \cite{PRLnoi,PREnoi,Langmuir}. Acquired images were analyzed through a custom-made program which first identifies the droplet contour and then fits it with a polynomial function, which is subsequently used to evaluate the front and rear contact points \cite{Toth11,Ferraro12}. After a transient time, the droplets acquire a stationary velocity that was evinced by the temporal evolution of the front contact point. For every solution, droplet sliding was observed at inclinations of the PC surface comprised between the sliding angle $\alpha _{c}$ (the smallest angle beyond which the droplet starts moving, see later) and the highest angle for which volume control was still accurate. \\

\begin{figure}[!htb]
\includegraphics[scale=1.06]{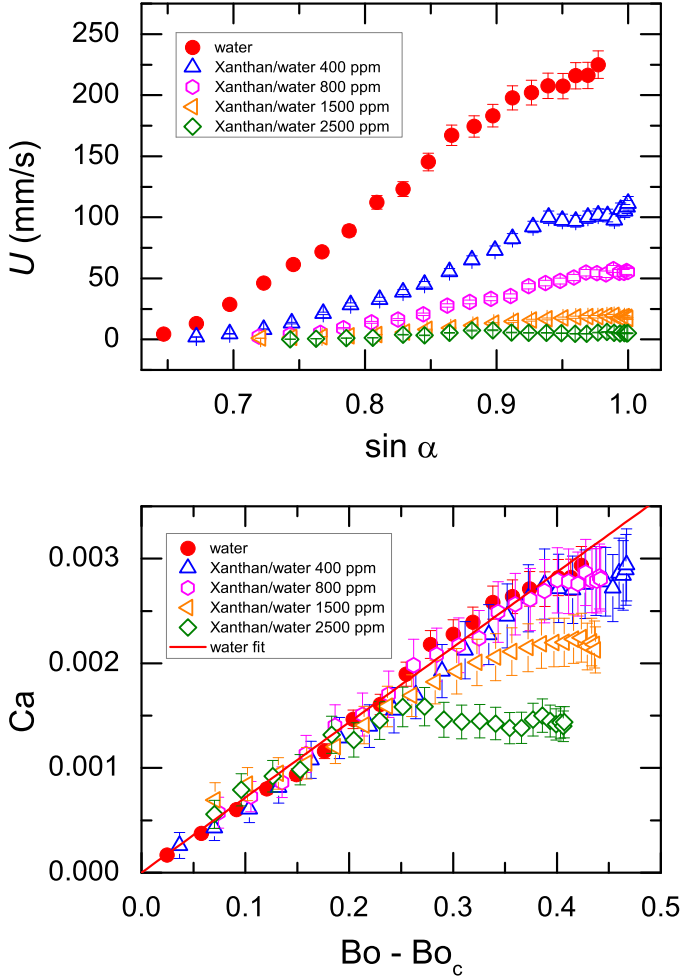}
\caption{Top panel: velocity of 30 $\protect\mu\text{L}$ droplets sliding down the homogeneous plane (PC) inclined by different angles $\protect\alpha$ for water and Xanthan/water solutions. Bottom panel: corresponding dimensionless $\mathrm{Ca}$ \textit{vs.} $\mathrm{{Bo}-{Bo}_{c}}$ trend. \label{fig:exp_xant}}
\end{figure}

The main experimental results, consisting in the dependence of the stationary droplet velocity $U$ on the inclination angle $\alpha $ of the PC surface, are shown in the top panel of figure \ref{fig:exp_xant}. By comparing the results of the polymeric solutions with those of water, we observe a decrease of $U$ as the polymer concentration increases due to the higher viscosity of the more concentrated solutions, which results in a smaller steady velocity at a fixed driving force \cite{Furmidge62,Podgorski01,Kimetal02,Rio05}. As a consequence, in order to better compare the behavior of the different liquids, the data have to be plotted in terms of dimensionless numbers which can be suggested by the analysis of the sliding of newtonian droplets \cite{Podgorski01,Kimetal02}, where the steady velocity $U$ comes out from the balance between the work done by the external driving force and the viscous dissipation. More quantitatively, a newtonian liquid droplet of volume $V$, density $\rho $ and viscosity $\eta $ sliding down an inclined plane tilted by an angle $\alpha $ is subject to the gravity force, the viscous drag and interfacial forces (see figure \ref{sketch}). The down-plane component of the droplet weight is $\rho gV\sin \alpha $, the viscous drag force is proportional to $\eta V^{1/3}U $ while the interfacial force is proportional to $\sigma V^{1/3}\Delta _{\theta }$, where $\sigma $ is the liquid-gas surface tension and $\Delta _{\theta }$ is a non dimensional factor depending on the perimeter-averaged projection of surface tension. In addition, the non vanishing difference between the advancing and the receding contact angles for small velocities leads to contact angle hysteresis~\cite{Furmidge62}. The critical inclination angle $\alpha_c$ is then identified as the angle below which the droplet stays pinned and it does not move. For inclination angles above $\alpha _{c}$, the balance of the gravity, viscous and capillary forces implies a linear scaling law~\cite{Podgorski01,Kimetal02} between the \textit{Capillary number} $\mathrm{{Ca}=\eta U/\sigma}$ and the \textit{Bond number} $\mathrm{{Bo}=V^{2/3}\rho g\sin \alpha /\sigma }$ 
\begin{equation}
\mathrm{{Ca}\propto {Bo}-{Bo_{c}}}  \label{eq:scaling}
\end{equation}%
where $\mathrm{{Bo}_{c}=V^{2/3}\rho g\sin \alpha _{c}/\sigma }$ depends on the wetting hysteresis through $\Delta _{\theta }$. 

For the non-newtonian Xanthan solutions the calculation of the Bond number is straightforward, on the other hand, due to the shear thinning behavior, the viscosity inside the droplet is obviously non constant and the computation of the Capillary number requires the evaluation of the viscosity corresponding to the sliding velocity $U$: one needs to devise a method to convert the data of viscosity versus shear-rate in viscosity versus $U$. For this practical purpose, we define an ``effective'' viscosity in terms of an effective shear rate $\dot{\gamma}_{eff}=U/\lambda$ 
\begin{equation}\label{viscoeff}
\eta_{eff}(U)=a \dot{\gamma}_{eff}^{-b}=a\left( \frac{U}{\lambda }\right)^{-b}
\end{equation}
where $a$ and $b$ are the parameters reported in Table \ref{viscosity_table}. The definition of this effective viscosity hinges on the introduction of the characteristic scale $\lambda$. The viscosity $\eta_{eff}$ in \eqref{viscoeff} is then used to calculate the Ca for the Xanthan droplets and the corresponding values are plotted as a function of $\mathrm{{Bo}-{Bo}_{c}}$ in the bottom graph of Figure \ref{fig:exp_xant}. Remarkably, a constant value of $\lambda $ ($\sim$ 10 $\mu $m) for all the Xanthan concentrations investigated, allows to obtain a good collapse of the non-newtonian data on the newtonian ones for small Bo. The scale $\lambda$ can be thought of as defining an effective gradient $U/\lambda$ which is ``representative'' for viscous dissipation inside the droplet. Such scale is indeed smaller than the macroscopic length scale (i.e. the capillary length) of the order of 1 mm and larger than the microscopic length (i.e. the slip length) ranging from 1 to $10^2$ nm \cite{Yueetal12}. However, at larger $\mathrm{{Bo}-{Bo}_{c}}$, non-newtonian data systematically deviate from the reference newtonian data, and even a plateau is emerging for the largest concentrations considered. We think there may be different reasons for these deviations. First, the characteristic scale $\lambda$ may not be constant as a function of the velocity: the non constancy of the viscosity inside the droplet may well induce a variation of $\lambda$ for larger velocities, with $\lambda$ being an increasing function of the velocity. To shed some lights on this issue, we have carried out some numerical simulations with a power-law fluid and we have not found evidences of this velocity dependence in the characteristic scale $\lambda$. These results are described in section \ref{sec:Simulations1}. We could, instead, relate the deviations from non-newtonian data to the presence of normal stresses inside the droplet, which are more important at high shear rates \cite{Rafai04,Macosko78,Helmreich95,BonnMeunier97,Zirnsak99,Lopez02,Stokes11,Gupta86,Jones87,Binnington87,Chai90,Bewersdorff88,Won04}, i.e. at high velocities and $\mathrm{Bo}$, while it is weaker at small inclinations where the behavior is similar to a newtonian fluid. With the measured velocities and the used values for $\lambda$ (10 $\mu $m), we indeed obtain $\dot{\gamma}_{eff} \sim 10^2 - 10^3$ s$^{-1}$ for the largest velocities of the more concentrated Xanthan solutions analyzed. This itself points to a representative value for the shear rate at which measured normal stresses are different from zero \cite{Rafai04,Macosko78,Helmreich95,BonnMeunier97,Zirnsak99,Lopez02,Stokes11,Gupta86,Jones87,Binnington87,Chai90,Bewersdorff88,Won04}, as discussed for the bottom panel of figure \ref{fig:ReoXanthan}.  An elastic non-viscous effect has been observed also in the flow of Xanthan solutions through a porous medium and has been ascribed to the presence of normal stresses \cite{Helmreich95}.\\
To shed some lights on this issue, we have carried out some numerical simulations with a normal-stress fluid where thinning effects are extremely reduced. These results are described in section \ref{sec:Simulations2}.

\section{Numerical Simulations}\label{sec:Simulations}

In our numerical study, we considered the diffuse-interface Navier-Stokes (NS) hydrodynamic equations of a binary mixture of two components~\cite{Yueetal12,Yue04,Yueetal05,Yueetal06a,Yueetal06b,Yueetal08} obtained with a lattice Boltzmann model (LBM) \cite{PREnoi} to simulate a droplet ($d$) with dynamic viscosity $\eta_{d}$ and density $\rho_{d}$ sliding in an outer ($o$) fluid with dynamic viscosity $\eta_{o}$ and density $\rho_{o}$~\cite{PREnoi,Langmuir}. We analyzed the case of a two-dimensional (cylindrical) droplet \cite{Moradi} with radius $R$ on a homogeneous surface with neutral wettability. The use of two-dimensional numerical simulations allowed us to better resolve the hydrodynamics inside the droplet. All the technical details for the simulation of newtonian droplets sliding down homogeneous surfaces are well described in our previous publication~\cite{PREnoi}. A few words are in order for the choice of the viscous ratio $\chi$ between the droplet and outer phases: approaching the extremely large experimental liquid-air viscous ratio is not easily achievable with the numerical models. One can partially cope with this problem by making the body force monotonously decrease to zero across the fluid-fluid interface, accounting for the fact that the outer phase remains inert in the limit of zero droplet size~\cite{Langmuir}. In fact, when working with very large viscosity ratios, numerical errors increase close to the interface \cite{Yueetal12}, especially close to the contact line. Fortunately, these deficiencies do not affect conclusions with viscous ratios~\cite{PREnoi,Yueetal12} up to a few tens. 

\subsection{Shear Thinning Fluid}\label{sec:Simulations1}

\begin{figure}[h!]
\includegraphics[scale=1.06]{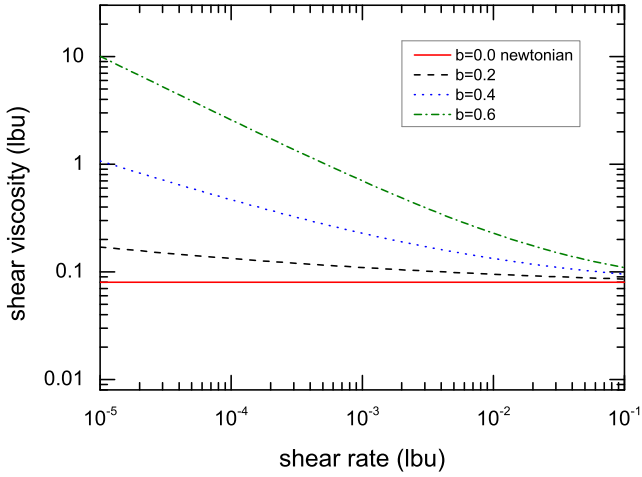}
\caption{Viscosity of the power-law fluid used in the numerical simulations. The exponent $b$ regulates the degree of thinning. On average, dissipation is larger at increasing $b$, in qualitative agreement with the behavior of the viscosity of Xanthan solutions at changing the polymer concentration.}
\label{fig:0}
\end{figure}

The numerical method, based on LBM, offers a unique advantage to attack these problems, being able to integrate simultaneously the non-trivial interface dynamics (due to the diffuse interface nature of the method) and the non-newtonian nature of the bulk phases, by making the relaxation times in LBM dependent on the local shear properties \cite{Gabbanelli}.  The equations we solve in the outer (o) and droplet (d) phases are the NS equations 
\be\label{NSc}
\begin{split}
\rho_{o,d} & \left[ \partial_t \bm u_{o,d} + ({\bm u}_{o,d} \cdot {\bm \nabla}) \bm u_{o,d} \right]  =  - {\bm \nabla}P_{o,d}+\\ &  {\bm \nabla} \left(\eta_{o,d} ({\bm \nabla} {\bm u}_{o,d}+({\bm \nabla} {\bm u}_{o,d})^{T})\right).
\end{split}
\ee
Here, ${\bm u}_{o,d}$ and $\eta_{o,d}$ are the velocity and the dynamic viscosity of the outer and droplet phase, respectively. The corresponding densities and pressures are indicated with $\rho_{o,d}$ and $P_{o,d}$, while $({\bm \nabla} {\bm u}_{o,d})^T$ indicates the transpose of $({\bm \nabla} {\bm u}_{o,d})$. As plotted in Figure \ref{fig:0}, the viscosity in the droplet phase is chosen to be
\be\label{eq:visco}
\eta_d(\dot{\gamma})=\begin{cases} \eta_0 +a \dot{\gamma}_0^{-b} \hspace{.2in} \dot{\gamma} \le \dot{\gamma}_0 \\ \eta_0+a \dot{\gamma}^{-b} \hspace{.2in} \dot{\gamma} > \dot{\gamma}_0 \end{cases}
\ee
where $a=0.01$ (lbu, lattice Boltzmann units) and where the cut-off shear rate $\dot{\gamma}_0=10^{-5}$ lbu is introduced to prevent a divergence of the viscosity at small shear rates. The background viscosity $\eta_0=0.07$ is chosen to avoid a zero viscosity in the high shear rate limit, which would cause numerical instabilities. For all practical purposes, in the range of shear rates relevant for our problem (see figure \ref{fig:0}), we are actually simulating a power-law fluid. In the outer matrix phase the viscosity is kept constant to $\eta_0=0.07$ so that the viscous ratio between the droplet and the outer phase ranges in the interval $[1-10]$ for the results presented.

\begin{figure}[t!]
\includegraphics[scale=1.06]{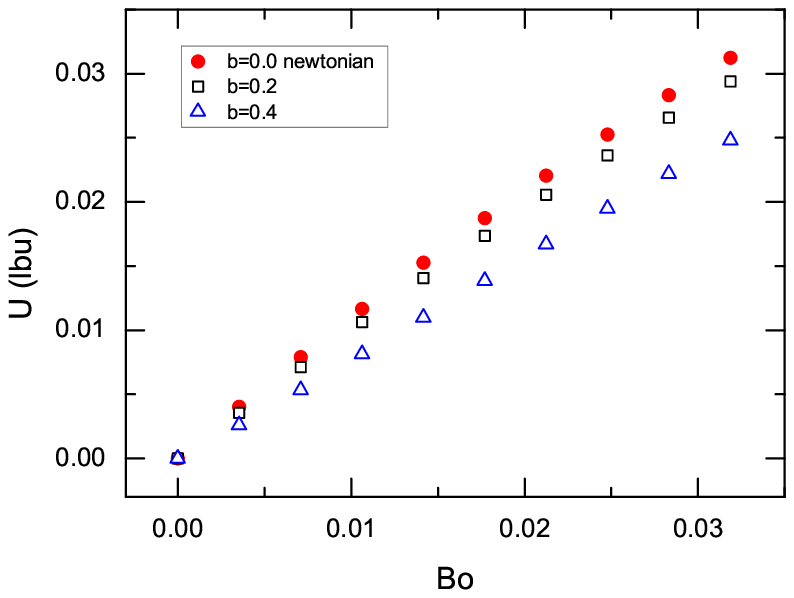}\\
\includegraphics[scale=1.06]{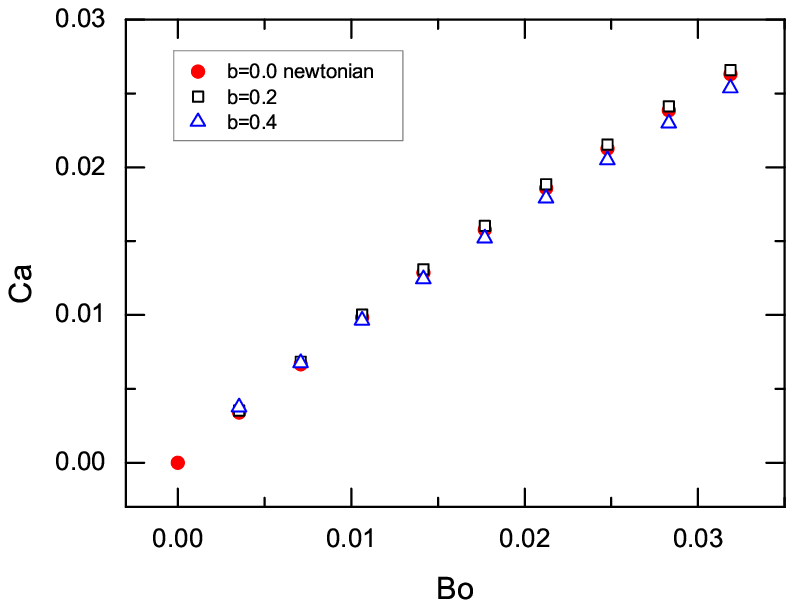}
\caption{Top panel: stationary velocity for sliding droplets at changing the thinning exponent $b$ in \eqref{eq:visco}. Bottom panel: rescaled data based on the lengthscale $\lambda=0.3$ (lbu) in the definition of the shear rate $\dot{\gamma}=U/\lambda$ used in the viscosity to identify the ``effective'' viscosity of the thinning fluid. The rescaling is chosen in such a way that the non-newtonian data overlap the newtonian data for small Bond numbers. Good collapse is also found at larger Bond number.\label{fig:1}}
\end{figure}

In the top panel of figure \ref{fig:1} we report the stationary velocity measurement for 3 cases: a newtonian case ($b=0$) and two viscoelastic cases with $b=0.2$ and $b=0.4$. Consistently with the fact that the viscosity chosen \eqref{eq:visco} produces a more viscous fluid at increasing $b$, the stationary droplet velocity is smaller at increasing the thinning exponent. However, by following the same procedure explained for the experimental data, once we tune the scale $\lambda$ to achieve good collapse on the newtonian data for small velocities, a very good collapse is obtained also for higher Ca, as shown in the bottom graph of Figure \ref{fig:1}. This surely indicates robustness in this ``practical'' procedure. It is true that some tiny deviations emerge for the largest Ca, but they are really very small and surely not at all reminiscent of the plateau behavior observed for the experimental data on Xanthan. 

\subsection{Normal Stress Fluid}\label{sec:Simulations2}

\begin{figure}[h!]
\includegraphics[scale=1.06]{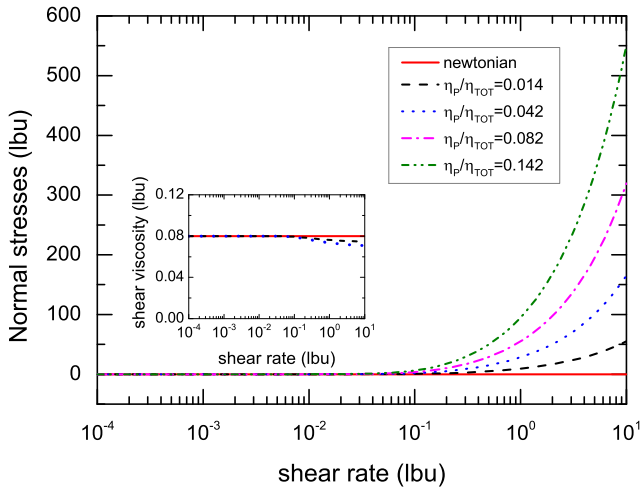}
\caption{Normal Stresses (main panel) and shear viscosity (inset) curves for the chosen viscoelastic solution. With the parameters chosen, the thinning effects are minimized while normal stresses are present for the range of shear rates relevant for our problem.}
\label{fig:2}
\end{figure}

In order to introduce normal stresses in the droplet phase, we used an available version of our code implementing the FENE-P constitutive model, obtained via a pre-averaging approximation applied to a suspension of non interacting Finitely Extensible Nonlinear Elastic (FENE) dumbbells. This is well-adapted for dilute (and semidilute) polymeric solutions and was previously used to analyze filament thinning of polymeric fluids in macroscopic experiments~\cite{Lindner03,Wagner05} and droplet breakup processes in microchannels~\cite{Arratia}. FENE-P fluids exhibit normal stress effects with a quadratic dependence on the shear rate. Such dependence on the shear rate contrasts with the sub-linear scaling law found for rigid polysaccharide solutions like Xanthan \cite{Stokes11}; in fact, we rather use this numerical model to ``generically'' highlight the effect of Normal Stresses. The algorithm we use is based on a hybrid combination of LBM with finite difference schemes, as detailed in \cite{SbragagliaGuptaScagliarini,SbragagliaGupta}. The equations that we solve in the droplet phase are the Navier-Stokes (NS) equations coupled to the FENE-P constitutive equations
\be\label{NSc}
\begin{split}
\rho_{d} & \left[ \partial_t \bm u_{d} + ({\bm u}_{d} \cdot {\bm \nabla}) \bm u_{d} \right]  =  - {\bm \nabla} P_{d}+\\ &  {\bm \nabla} \left(\eta_{d} ({\bm \nabla} {\bm u}_{d}+({\bm \nabla} {\bm u}_{d})^{T})\right)  +\frac{\eta_{P}}{\tau_{P}}{\bm \nabla} \cdot [f(r_{P}){\bm {\bm {C}}}].
\end{split}
\ee
\be\label{FENEP}
\begin{split}
\partial_t {\bm {C}} + (\bm u_{d} \cdot {\bm \nabla}) {\bm {C}}  =  {\bm {C}} \cdot ({\bm \nabla} {\bm u}_{d}) + & {({\bm \nabla} {\bm u}_{d})^{T}} \cdot {\bm {C}}  \\  -& \left(\frac{{f(r_{P}){\bm {C}} }- {{\bm I}}}{\tau_{P}}\right).
\end{split}
\ee
where $\eta_{P}$ is the viscosity parameter for the FENE-P solute, $\tau_P$ the polymer relaxation time,  ${\bm {C}}$ the polymer-conformation tensor, ${\bm I}$ the identity tensor, $f(r_P)\equiv{(L^2 -3)/(L^2 - r_P^2)}$ the FENE-P potential that ensures finite extensibility, $ r_P \equiv \sqrt{Tr({\bm {C}})}$ and $L$ is the maximum possible extension of the polymers~\cite{bird87,Herrchen97}. In the outer phase we just consider the NS equations with viscosity $\eta_P+\eta_{d}$, i.e. with the same shear viscosity as the droplet phase. The chosen value of $L^2$ is so large that the system only displays very weak thinning effects, while possessing normal stresses increasing with the squared shear rate (see figure \ref{fig:2}). The value of the polymer relaxation time $\tau_P$ is chosen in such a way that the range of sliding velocity produces shear rate in the wedge which are of the order of magnitude of $\tau_P^{-1}$. We expect that this choice allows us to appreciate the effects of the normal stresses on the Capillary-Bond curve. Further details on the algorithm used can be found in \cite{SbragagliaGuptaScagliarini,SbragagliaGupta}.

\begin{figure}[h!]
\includegraphics[scale=1.06]{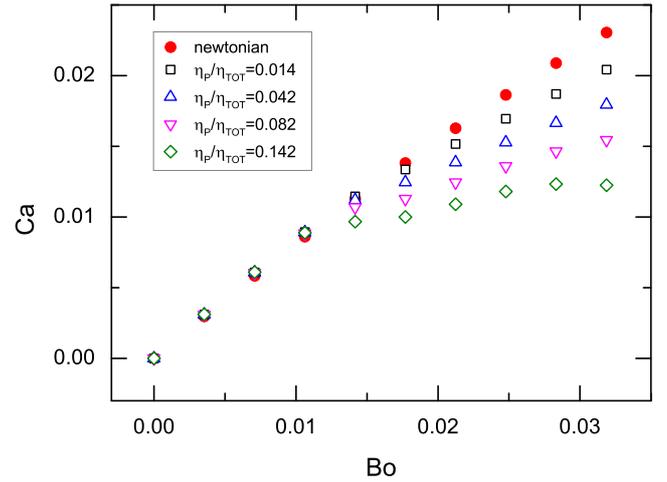}
\caption{Capillary {\it vs.} Bond number curve for viscoelastic fluid with rheological properties described in figure \ref{fig:2}. The concentration $\eta_P/\eta_{TOT}=\eta_P/(\eta_d+\eta_P)$ is varied from $0.01$ to $0.14$.}
\label{fig:3}
\end{figure}

Results for the Ca {\it vs.} Bo curve are presented in figure \ref{fig:3}. The concentration $\eta_P/\eta_{TOT}=\eta_P/(\eta_d+\eta_P)$ is systematically varied from $0.01$ to $0.15$. The viscosity ratio of the two fluids was chosen in such a way that $\chi =(\eta _{d}+\eta _{P})/\eta_o=1.0$. This allowed to study the properties of viscoelastic droplets ($\eta _{P}\neq 0$) and compare them with the corresponding newtonian ($\eta _{P}=0$) case. It is readily verified that the deviations from linearity first emerge with a slight sublinear behavior for the smallest concentrations, while they produce a plateau for the largest concentrations. The behavior observed in the presence of normal stresses points to the fact that effects other than viscous dissipation come into play to balance the work done by the external driving. We have verified in the numerical simulations that normal stresses provide an additional driving and accelerate the flow near the contact line \cite{Rafai04}, and the effect does so more strongly as the shear rate is increased. Parts closest to the rear wedge move faster and there the polymer feedback stress is enhanced to provide a counter force against sliding. It is interesting to observe that the characteristic Ca at which the linear behavior starts to be violated is essentially unchanged. Preliminary numerical simulations show that such ``critical'' Ca depends on the relaxation time $\tau_P$: numerical results obtained at changing the relaxation time $\tau_P$ will be the object of a forthcoming publication, which will be useful to better characterize the relative importance of both the polymer relaxation time and the polymer concentration in generating sublinear and plateau behaviours.

\section{Conclusions}\label{sec:conclusion}

We have investigated the sliding of droplets made of polysaccharides in aqueous solutions. In particular, we used Xanthan, a stiff rodlike polysaccharide exhibiting a non-newtonian behavior, notably characterized by a shear-rate dependence of the viscosity. The sliding of non-newtonian droplets was compared to the one of water through the relation between the Capillary number (i.e. the dimensionless velocity) and the Bo number (i.e. the dimensionless volume force), as the driving force is changed by varying the plane inclination. To account for the spatially varying viscosity inside the droplet, we argued that an effective viscosity could be defined to quantitatively compare with the corresponding newtonian problem.  The usual linear behavior of newtonian solutions is found to well adapt to the non-newtonian solutions only up to a given Bo number, above which deviations emerge, as the linear behavior is first replaced by a sublinear behavior (small concentrations) and then by a more visible plateau exhibiting a saturation in the Capillary number. To discriminate to what degree the observed behavior could be due to the thinning effects or other non-newtonian effects (e.g. Normal Stresses), results from experiments were complemented with lattice Boltzmann numerical simulations of sliding droplets, aimed to disentangle the influence that non-newtonian flow properties have on the sliding. Numerical simulations reveal that the observed behavior is more likely attributed to the emergence of normal stresses inside the non-newtonian droplet, rather than thinning effects. \\
For future studies it is definitively warranted a systematic investigation of the sliding problem for other polymeric solutions. This is the case, for example, of polyacrylamide solutions \cite{Rafai04}, where the viscosity is known to be only weakly dependent on the shear rate while being characterized by normal stress effects. Work in this direction is ongoing.  \\

The authors kindly acknowledge funding from the European Research Council under the European Community's Seventh Framework Programme (FP7/2007-2013)/ERC Grant Agreement No. 279004. We are particularly grateful to Prof. Giovanni Lucchetta (Department of Industrial Engineering, University of Padova, Italy) for fruitful discussions and support about the rheological characterizations.

\bibliographystyle{spphys}       

\bibliography{sample}

\end{document}